# Resilience-Oriented DG Siting and Sizing Considering Energy Equity Constraint

Chenchen Li, *Graduate Student Member, IEEE*, Fangxing Li, *Fellow, IEEE*, Sufan Jiang, *Member, IEEE*, Jin Zhao, *Member, IEEE*, Shiyuan Fan, *Member, IEEE*, Leon M. Tolbert, *Fellow, IEEE*

*Abstract*—Extreme weather events can cause widespread power outages and huge economic losses. Low-income customers are more vulnerable to power outages because they live in areas with poorly equipped distribution systems. However, existing approaches to improve grid resilience focus on the overall condition of the system and ignore the outage experiences of low-income customers, which leads to significant energy inequities in resilience. Therefore, this paper explores a new resilience-oriented planning method for distributed generator (DG) siting and sizing, by embedding an additional energy equity constraint (EEC). First, the expected load shedding index (ELSI) is defined as the ratio of the load shedding to the original load, which quantifies the resilience-oriented energy equity. Then, the DG siting and sizing problem is formulated as a two-stage stochastic programming with the EEC. The first stage determines the optimal sites and sizes of DG units under investment constraints and EECs, while the second stage optimizes expected costs of unserved load. A subsidiary variable is introduced to ensure the model's solvability. Finally, numerical studies are performed on the IEEE 33-bus and 123-bus systems to verify the effectiveness of the proposed DG planning model in achieving energy equity. Three observations are presented as future guidelines for resilience-oriented DG planning.

## Nomenclature

*Indices and sets*

| | |
|---|---|
| $s, \Omega_S$ | Index and set of scenarios |
| $t, \Omega_T$ | Index and set of time intervals |
| $i, j, \Omega_N$ | Index and set of buses |
| $\Omega_G$ | Set of distributed generators (DGs) |
| $(i,j), \Omega_L$ | Index and set of lines |
| $\Omega_I^{t,s}$ | Set of buses of island at time $t$ in scenario $s$ |
| $\Omega_{IA}^{t,s}$ | Set of buses of island with energy source at time $t$ in scenario $s$ |
| $\Omega_{IB}^{t,s}$ | Set of buses of island without energy source at time $t$ in scenario $s$ |
| $\Omega_{LF}^{t,s}$ | Set of faulted lines at time $t$ in scenario $s$ |
| $\Omega_{SVC}$ | Set of static var compensators (SVCs) |
| $\Omega_{NLH}$ | Set of high- and medium-income communities |
| $\Omega_{LH}$ | Set of low-income communities |

*Parameters*

| | |
|---|---|
| $\rho_s$ | Probability of scenario $s$ |
| $P_{i,t,s}^D/Q_{i,t,s}^D$ | Real/reactive load demand of bus $i$ at time $t$ in scenario $s$ |
| $N_{DG}$ | Maximum number of DGs to be installed |
| $C_{DG}$ | Investment budget for the DG installation |
| $P^{max}$ | Maximum rated power of DGs to be installed |
| $e_i^*$ | Energy equity criterion of community $i$ |
| $E$ | Set of energy equity criterion of all communities |
| $\beta$ | Payment for load shedding |
| $N$ | Number of buses in the distribution system |
| $\theta(j)$ | Set of child buses of bus $j$ |
| $\pi(j)$ | Set of parent buses of bus $j$ |
| $P_j^{G,min}/Q_j^{G,min}$ | Minimum real/reactive power of DG $j$ |
| $Q_j^{C,min}/Q_j^{C,max}$ | Minimum/maximum reactive power of SVC $j$ |
| $\psi_j$ | Power factor of DG $j$ |
| $V^{min}/V^{max}$ | Minimum/maximum bus voltage |
| $R_{ij}/X_{ij}$ | Reactance/inductance of line $(i,j)$ |
| $V_{sub,t,s}$ | Bus voltage of the substation at time $t$ in scenario $s$ |
| $S_{ij}^{max}$ | Thermal limit of line $(i,j)$ |

*Variables*

| | |
|---|---|
| $e_i$ | Expected load shedding index of community $i$ |
| $\Delta P_{i,t,s}/\Delta Q_{i,t,s}$ | Real/reactive load shedding of bus $i$ at time $t$ in scenario $s$ |
| $k_i$ | Binary variable indicating whether a DG is installed at bus $i$ |
| $P_i^R$ | Rated power of DG installed at bus $i$ |
| $z_{ij,t,s}$ | Binary variable indicating line switching status in scenario $s$ (1=closed, 0=open) |
| $N_{t,s}^I$ | Number of buses in island at time $t$ in scenario $s$ |
| $F_{ij,t,s}$ | Power flow of line $(i,j)$ in the virtual network at time $t$ in scenario $s$ |
| $P_{j,t,s}^G/Q_{j,t,s}^G$ | Real/reactive power of DG on bus $j$ at time $t$ in scenario $s$ |
| $Q_{j,t,s}^C$ | Reactive power of SVC on bus $j$ at time $t$ in scenario $s$ |
| $V_{j,t,s}$ | Bus voltage of bus $j$ at time $t$ in scenario $s$ |
| $P_{ij,t,s}/Q_{ij,t,s}$ | Real/reactive power flow of line $(i,j)$ at time $t$ in scenario $s$ |
| $P_{j,t,s}/Q_{j,t,s}$ | Real/reactive power injection of bus $j$ at time $t$ |

This work was supported in part by the U.S. DOE Project titled "Equitable, Affordable & Resilient Nationwide Energy System Transition (EARNEST)."

C. Li, F. Li, S. Jiang, S. Fan, L. M. Tolbert are with Department of Electrical Engineering and Computer Science, The University of Tennessee, Knoxville, TN, 37996, USA.

J. Zhao is with Department of Electronic and Electrical Engineering, School of Engineering, Trinity College Dublin, Dublin, Ireland, D02PN40.

Corresponding author: F. Li (email: fli6@utk.edu).

in scenario *s*

## I. Introduction

POWER system resilience is defined as the ability of a power system to withstand and recover from high-impact low-probability hazards like hurricanes, tornadoes, and winter storms [1]. These extreme weather events have resulted in significant power outages and economic losses in recent years [2]. For example, Hurricane Maria in 2017 and Hurricane Fiona in 2022 caused severe damage to the power grid in Puerto Rico, plunging the entire island into darkness and resulting in an estimated damage cost of $117 billion [3].

Affected by severe winter storms in 2021, at least 146 people died in Texas from hypothermia, and the cost of the multi-day power outage was estimated at $195 billion [4, 5]. Additionally, service areas with more minority populations were four times more likely to lose power than white-majority neighborhoods in Texas [6]. Blacks, Hispanics, and Asians accounted for 72% of carbon monoxide poisoning cases using unconventional heat sources [7]. This phenomenon shows that devastating effects of power outages are not equally distributed.

This unequal resilience of power systems can also be found in other cases. In California, the distribution system service areas where low-income communities are located have a safety deficiency during wildfires [8]. According to tax appraisal data, low-income areas suffered more damage during Hurricane Andrew and Hurricane Ike and recovered more slowly after these hurricanes [9]. Additionally, customers with social vulnerabilities experienced longer power outages during Hurricane Irma [10]. During the early period of the COVID-19 pandemic, it was estimated that Hispanic households were 4.7 times more likely than White households to be disconnected from the grid, and after 12 months were 2.4 times more likely to be disconnected [11]. The above phenomena reflect that although a natural catastrophe is bad for everyone, it is even worse for low-income communities.

Much research has studied planning models to increase the resilience of the power system. Usually, the cost of total unserved loads is minimized by the planning model, improving the resilience of the whole system [12]. However, the individual customers' outage situations are not considered, thus resulting in the resilience inequity described previously. At present, increasing research is devoted to addressing this resilience inequity in power systems. It is stated that equity should be considered in the resilience planning model and recovery process [13, 14]. The equity issues in power system resilience planning and operation are discussed, and a power grid resilience enhancement framework is proposed, which covers different stages of disaster management and different dimensions of energy equity [15]. Integrating equity into the initial planning model will alter the installation strategy for power sources, enhancing the resilience of specified customers and minimizing disparities in outage experiences among customers.

Research efforts have been undertaken to incorporate equity into planning models. For example, an optimization framework jointly considering equitable resiliency and resource utilization has been proposed to obtain the allocation strategy of electric vehicles' charging infrastructure [16]. Additionally, a dispatch model of mobile energy storage systems was developed to achieve an equitable dispatch strategy during power outages, in which the objective function minimizes the total energy not served weighted by social vulnerability of each node [17].

To address power outages caused by wildfires and decrease the probability of power lines catching fire, a cost allocation scheme with an income threshold was proposed to underground fire-prone lines. Under this scheme, the utility-wide cost of undergrounding fire-prone lines is borne by households with median income less than the threshold, and the local cost is borne by remaining households [8]. The Grid-Aware Tradeoff Analysis framework was presented to select the best backup power systems for expanded Resilience Hubs to enhance resilience, where the equity-weighted value provided by a backup power system is quantified by the equity-weighted outage mitigation performance [18].

Although the above models and methods succeed in certain cases, there are still challenges in integrating energy equity into the planning model, especially in the DG planning stage. Since there are growing DG penetrations at the distribution level, it is difficult to achieve equity while enhancing the resilience of a studied power system. First, there is a lack of a general metric for quantifying energy equity in resilience for the power system. Additionally, energy equity is a sociological problem, and it is difficult to quantitatively integrate this sociological concern into the initial technical problems. Thus, it is a challenge to obtain an adaptive, justified solution method for the planning model considering energy equity.

To address the above challenges, we propose a two-stage stochastic programming for distributed generator (DG) siting and sizing with the energy equity constraint (EEC) considered. The technical contributions of this work are summarized as follows:

- A mathematical model is proposed to quantify energy equity in resilience for a power system, defined as the expected load shedding index (ELSI), which focuses on the outage performance of individual community.
- A two-stage stochastic programming is proposed for the DG siting and sizing problem, where energy equity is considered as constraints. In this mathematical model, the first stage aims to minimize the expected costs of unserved load under the investment constraints and the EEC. The second stage optimizes the cost of unserved load in each scenario with a known installation strategy of DG units.
- To solve the proposed optimization model, a subsidiary variable is introduced into the EEC. This conversion ensures the stochastic mixed-integer linear programming is solvable.

The remaining part of the paper is organized as follows. Section II proposes the two-stage stochastic programming for DG siting and sizing with EECs. Section III describes the stochastic scenarios generation process and reduction method. Section IV proposes the solution method. Section V presents numerical results of the IEEE 33-bus system and 123-bus system. Finally, Section VI concludes this paper.

## II. TWO-STAGE STOCHASTIC PROGRAMMING FOR DG SITING AND SIZING WITH ENERGY EQUITY CONSTRAINT

Resilience-oriented DG siting and sizing problem with the EEC is formulated as a two-stage stochastic programming, which is presented in this section. The overall framework is illustrated in Fig. 1.

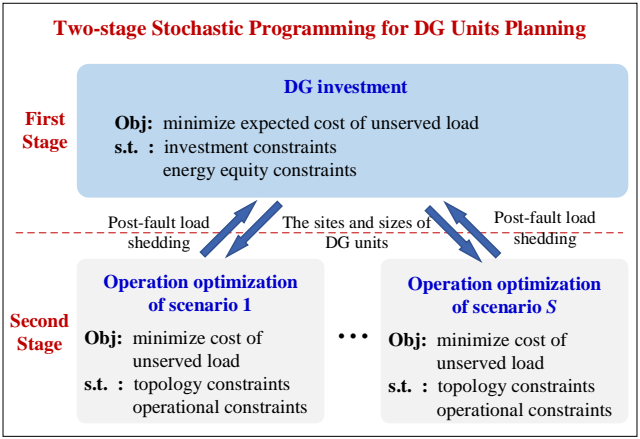

Fig. 1. Framework of the proposed two-stage stochastic programming.

### A. Quantitative Model of Energy Equity

Load shedding varies among different communities under extreme weather. Low-income communities are often at a higher risk of power outages due to poor infrastructure in these communities [8, 19]. In this perspective, the ELSI is proposed to quantify energy equity in this paper. This simple yet effective index is defined as the ratio of the load shedding to the original load. The mathematical model is shown as follows.

$$e_i = \sum_{s \in \Omega_S} \rho_s \cdot \sum_{t \in \Omega_T} \frac{\Delta P_{i,t,s}}{P_{i,t,s}^D}, \forall i \in \Omega_N \quad (1)$$

### B. The First Stage Problem

In the proposed two-stage model for resilience-oriented DG planning considering the EEC, the first stage minimizes the expected costs of unserved load while satisfying both the investment constraints of DG and the EEC. The decision variables of the first stage are the optimal sites and sizes of DG units, i.e., $x = (k_i, P_i^R), i \in \Omega_N$.

$$\min \mathbb{E}[g(x,s)] \quad (2)$$

$$s.t. \ 0 \leq \sum_{i \in \Omega_N} k_i \leq N_{DG} \quad (3)$$

$$\sum_{i \in \Omega_N} k_i \cdot \alpha^p \cdot P_i^R + \alpha^e \cdot k_i \leq C_{DG} \quad (4)$$

$$0 \leq P_i^R \leq k_i \cdot P^{\max}, \forall i \in \Omega_G \quad (5)$$

$$\mathbb{E}[g(x,s)] = \sum_{s \in \Omega_S} \rho_s \cdot g(x,s) \quad (6)$$

$$e_i \leq e_i^*, \forall i \in \Omega_N \quad (7)$$

where (3) is the constraint on the number of DG units installed. Equation (4) limits the investment budget of DG units, which is formulated as a linear function consisting of variable and fixed costs. $\alpha^p$ and $\alpha^e$ are the constant parameters related to investment cost [12, 20]. Equation (5) is the DG unit capacity constraint, associated with the binary variable $k_i$. If there is a DG unit installed at the bus $i$ (i.e., $k_i = 1$), the DG unit's rated output can be between 0 and its maximum rated power $P^{max}$; otherwise, the rated output is 0. Equation (6) shows that the expected cost of the second stage is the weighted-sum cost of stochastic scenarios. Equation (7) represents the EEC of communities located at each bus.

### C. The Second Stage Problem

The second stage aims to optimize the cost of unserved load in each scenario under the condition that the location and rated power of DG units are determined. The objective function of scenario $s$ is formulated as (8). In the studied distribution system, some distribution lines are equipped with automatic switches that can receive control signals from the system operator. Therefore, topology reconfiguration can reduce the load shedding after the outage. Further, many DG units can work normally in extreme weather because they are typically placed indoors or placed outdoors with shelters, making them less susceptible to hurricane damage [12]. Through rescheduling the operation of installed DG units after an electrical fault, electrical islands can be formed to restore power to some communities experiencing outages.

$$g(x,s) = \min \beta \cdot \sum_{t \in \Omega_T} \Delta t \cdot \sum_{j \in \Omega_N} \Delta P_{j,t,s} \quad (8)$$

*1) Topology Constraints:* A distribution system can be regarded as a directed graph consisting of buses and branches. Normally, there is a sectionalizing switch or tie switch in the branch, as shown in Fig. 2. According to graph theory, a radial graph is defined as a connected graph without loops, consisting of $n$-1 lines for a graph with $n$ buses [21]. To model topology constraints, the single commodity flow is applied in this paper [12]. The principle is building a lossless fictitious system that has the same topology and switch status variables as the original system. Moreover, it is assumed that the load demand of a non-source bus is 1.0 in the virtual system. Then, the radial topology constraints are formulated as following linear equations.

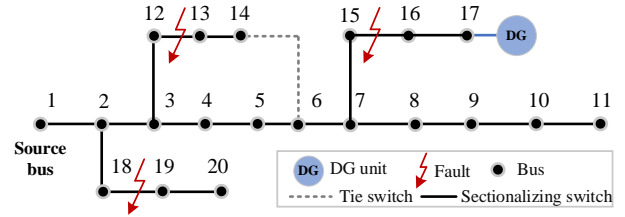

Fig. 2. Example of a distribution system.

$$\sum_{(i,j) \in \Omega_L} z_{ij,t,s} = N - 1 - N_{t,s}^I \quad (9)$$

$$\begin{cases} z_{ij,t,s} = 0, (i,j) \in \Omega_{LF}^s \\ z_{ij,t,s} = 1 \text{ or } 0, (i,j) \in \Omega_L \setminus \Omega_{LF}^s \end{cases} \quad (10)$$

$$\begin{cases} \sum_{y\in\theta(j)} F_{jy,t,s} - \sum_{i\in\pi(j)} F_{ij,t,s} = -1, \forall j \in \Omega_N, j \neq 1, j \notin \Omega_{IB}^{t,s} \\ \sum_{y\in\theta(j)} F_{jy,t,s} - \sum_{i\in\pi(j)} F_{ij,t,s} = 0, \forall j \in \Omega_{IB}^{t,s} \\ \sum_{y\in\theta(j)} F_{jy,t,s} = N - 1 - N_{t,s}^I, j = 1 \end{cases} \quad (11)$$

$$\Omega_I^{t,s} = \{\Omega_{IA}^{t,s}, \Omega_{IB}^{t,s}\} \quad (12)$$

$$-M \cdot z_{ij,t,s} \leq F_{ij,t,s} \leq M \cdot z_{ij,t,s}, (i,j) \in \Omega_L \quad (13)$$

where (9) shows that the number of lines with closed switches equals the number of non-source buses minus the number of islanded buses. The islanded bus means there is no path connecting this bus and the source bus even if closing all tie switches. For example, if line (18, 19) in Fig. 2 is tripped, buses 19 and 20 cannot be connected to the source bus. If line (12, 13) is tripped, buses 13 and 14 can be connected to the source bus by closing the tie switch at line (6, 14). Equation (10) represents that switches of faulted lines are open, switches of other lines are open or closed, depending on the optimal reconfiguration strategy. The power flow balance is guaranteed by (11). Equation (12) shows there are two types of islands, i.e., islands with DG units (e.g., $\Omega_{IA}^{t,s} = \{16,17\}$ in Fig. 2) and islands without DG units (e.g., $\Omega_{IB}^{t,s} = \{19,20\}$ in Fig. 2). Equation (13) is the line flow constraint, where $M$ is a big number.

*2) Operating constraints:* Optimizing the cost of unserved load should also satisfy DG units' operating constraints and power flow constraints, given by (14)-(26).

$$P_j^{G,\min} \leq P_{j,t,s}^G \leq P_j^R, \forall j \in \Omega_G \quad (14)$$

$$Q_j^{G,\min} \leq Q_{j,t,s}^G \leq P_{j,t,s}^G \cdot \tan(arc\cos\psi_j), \forall j \in \Omega_G \quad (15)$$

$$Q_j^{C,\min} \leq Q_{j,t,s}^C \leq Q_j^{C,\max}, \forall j \in \Omega_{SVC} \quad (16)$$

$$V^{\min} \leq V_{j,t,s} \leq V^{\max}, \forall j \in \Omega_N \quad (17)$$

$$-M \cdot (1 - z_{ij,t,s}) \leq V_{i,t,s} - V_{j,t,s} - \frac{R_{ij} \cdot P_{ij,t,s} + X_{ij} \cdot Q_{ij,t,s}}{V_0}$$
$$\leq M \cdot (1 - z_{ij,t,s}), \forall i, j \in \Omega_N, (i,j) \in \Omega_L \quad (18)$$

$$V_{1,t,s} = V_{sub,t,s} \quad V_{i,t,s} = V_0, \text{ if } k_i = 1 \& i \in \Omega_{IA}^{t,s} \quad (19)$$

$$-M \cdot z_{ij,t,s} \leq P_{ij,t,s} \leq M \cdot z_{ij,t,s}, (i,j) \in \Omega_L \quad (20)$$

$$-M \cdot z_{ij,t,s} \leq Q_{ij,t,s} \leq M \cdot z_{ij,t,s}, (i,j) \in \Omega_L \quad (21)$$

$$P_{j,t,s} = P_{j,t,s}^G - (P_{j,t,s}^D - \Delta P_{j,t,s}), \forall j \in \Omega_N \quad (22)$$

$$Q_{j,t,s} = Q_{j,t,s}^G + Q_{j,t,s}^C - (Q_{j,t,s}^D - \Delta Q_{j,t,s}), \forall j \in \Omega_N \quad (23)$$

$$P_{j,t,s} = \sum_{h\in\theta(i)} P_{jh,t,s} - \sum_{i\in\pi(j)} P_{ij,t,s}, \forall j \in \Omega_N, j \neq 1 \quad (24)$$

$$Q_{j,t,s} = \sum_{h\in\theta(i)} Q_{jh,t,s} - \sum_{i\in\pi(j)} Q_{ij,t,s}, \forall j \in \Omega_N, j \neq 1 \quad (25)$$

$$(P_{ij,t,s})^2 + (Q_{ij,t,s})^2 \leq (S_{ij}^{\max})^2, (i,j) \in \Omega_L \quad (26)$$

where (14) – (16) are the real and reactive power output constraints. Equation (17) ensures that the bus voltage is within the stable operating range. The voltage drop constraint is guaranteed by (18), and $V_0$ is the rated voltage. Equation (19) ensures that there is a slack bus with a reference voltage in a source-connected network. In the main radial network, bus 1 (substation bus) served as the slack bus. In the island with an energy source, the bus equipped with DG unit functions as the slack bus. Equations (20) and (21) show that the power flow through line $(i, j)$ is zero if a switch in this line is open. The real and reactive power injection to bus $j$ is shown in (22) and (23). The real and reactive power balance is ensured by (24) and (25), respectively. Equation (26) shows the thermal limits of the lines in the distribution system. To simplify the computation, the thermal limits are linearized as following three piecewise linear constraints (27) – (29) [12].

$$-2 \cdot S_{ij}^{\max} \leq \sqrt{3} \cdot P_{ij,t,s} + Q_{ij,t,s} \leq 2 \cdot S_{ij}^{\max}, (i,j) \in \Omega_L \quad (27)$$

$$-S_{ij}^{\max} \leq P_{ij,t,s} \leq S_{ij}^{\max}, (i,j) \in \Omega_L \quad (28)$$

$$-2 \cdot S_{ij}^{\max} \leq \sqrt{3} \cdot P_{ij,t,s} - Q_{ij,t,s} \leq 2 \cdot S_{ij}^{\max}, (i,j) \in \Omega_L \quad (29)$$

*D. The Compact Notation*

To make the model concise and clear, a compact notation is applied to represent the proposed two-stage stochastic programming. The first stage is as follows.

$$\min \sum_{s\in\Omega_S} \rho_s \cdot g(\boldsymbol{x}, s) \quad (30)$$

$$s.t. \quad \boldsymbol{Ax} \leq \boldsymbol{b} \quad (31)$$

$$\boldsymbol{d}^T \boldsymbol{y} \leq \boldsymbol{E} \quad (32)$$

where $\boldsymbol{x} \in \mathbb{Z}_+^{n_1} \times \mathbb{R}_+^{n_2-n_1}$ are the decision variables which include DG units' location and rated power. $n_1$ is the number of candidate buses for installing DG units, and $n_2$ is the number of total decision variables. $g(\boldsymbol{x}, s)$ is the function of the load shedding w.r.t decision variables and scenarios. $\boldsymbol{A} \in \mathbb{R}_+^{m_1 \times n_2}, \boldsymbol{b} \in \mathbb{R}_+^{m_1}$, $m_1$ is the number of constraints (3) – (5). $\boldsymbol{E} \in \mathbb{R}_+^N$, $N$ is the number of buses in the distribution system. $\boldsymbol{y} \in \mathbb{R}_+^N$ represents decision variables of the second stage, i.e., load shedding of each bus in scenarios.

The second stage is given by:

$$g(\boldsymbol{x}, s) = \min \boldsymbol{q}^T \boldsymbol{y} \quad (33)$$

$$s.t. \quad \boldsymbol{T}(s)\boldsymbol{x} + \boldsymbol{W}(s)\boldsymbol{y} \leq \boldsymbol{H}(s) \quad (34)$$

where $\boldsymbol{T} \in \mathbb{R}_+^{m_2 \times n_2}, \boldsymbol{W} \in \mathbb{R}_+^{m_2 \times N}, \boldsymbol{H} \in \mathbb{R}_+^{m_2}$, $m_2$ is the number of constraints (9) – (29).

III. STOCHASTIC SCENARIO GENERATION AND REDUCTION

The two-stage stochastic programming described in (30) – (34) is an optimization problem with stochastic scenarios. These scenarios account for variability in nodal load and the potential for line failures due to extreme weather events. It is infeasible and unnecessary to exhaust all scenarios for solving the two-stage stochastic programming. A widely used method for addressing this issue is sampling average approximation (SAA), whose principle is randomly selecting a finite number of scenarios [22]. This section details the scenario generation process based on the SAA principle. Additionally, a scenario reduction method using K-means clustering is employed to further simplify the problem, making it more tractable.

## A. Stochastic Scenarios Generation

In this paper, two uncertain factors are considered in the generation of stochastic scenarios.

*1) Uncertainty of Fault Location:* During extreme weather events, the performance of lines is mainly affected by weather and lines' conditions. Since a distribution system covers a small geographical area, its feeders and laterals are subjected to similar weather conditions [23]. However, due to differences in the condition of the distribution lines, the likelihood of faults varies for different lines. Specifically, lines in low-income communities tend to be in poorer condition and more prone to be tripped [8, 19]. Therefore, the probability of scenarios with tripped lines in low-income communities is higher than that of scenarios with tripped lines in high- or middle-income communities. For a better illustration, it is simply assumed that the probability of scenarios with tripped lines in high-income or middle-income communities is the same.

The objective function of the first stage model includes the sum of each term $\rho_s \cdot g(x,s)$. When there is only one line tripped, $g(x,s) = 0$ because all loads can be restored by network reconfiguration. Therefore, DG units are not needed, and we could neglect these scenarios with only one line tripped. Additionally, if there are four lines tripped simultaneously, the probability $\rho_s$ will be extremely small, then $\rho_s \cdot g(x,s)$ is small no matter how to allocate the DG units. Thus, these scenarios with four or more lines tripped do not need to be considered. Scenarios with two or three lines tripped are exhausted and the sum of their probabilities is 1.

*2) Uncertainty of Nodal Load:* We assume that there are the same normalized real and reactive load profiles for buses in the distribution system under each scenario [24]. Additionally, a random multiplier is applied to nodal loads to simulate the stochastic characteristics of the nodal load.

$$P_{j,t,s}^D = \tau_{j,t,s} \cdot M_{j,t}^P \cdot P_{t,s}^D, \forall j \in \Omega_N \quad (35)$$

$$Q_{j,t,s}^D = \tau_{j,t,s} \cdot M_{j,t}^Q \cdot Q_{t,s}^D, \forall j \in \Omega_N \quad (36)$$

where $\tau_{j,t,s}$ is the random multiplier of scenarios $s$, following a normal distribution, $\tau_{j,t,s} \sim N(1, 0.04^2)$ [24, 25]. $M_{j,t}^P$ and $M_{j,t}^Q$ are the normalized real and reactive load profiles. $P_{t,s}^D$ and $Q_{t,s}^D$ are the total real load and reactive load of the distribution system in scenario $s$.

## B. Scenario Reduction Based on K-means Clustering

Having too many scenarios decreases computational efficiency while having too few scenarios reduces computational accuracy. To achieve a trade-off between accuracy and efficiency, the scenario reduction method based on the K-means clustering algorithm is developed in this work to eliminate similar scenarios. The unserved energy $\triangle E_s$ in the distribution system can be obtained by solving the second-stage problem (33) – (34) for scenario $s$.

$$\triangle E_s = (\triangle E_{1,s}, \triangle E_{2,s}, \cdots, \triangle E_{N,s}), s \in \Omega_S \quad (37)$$

where $\triangle E_{j,s}$ is the accumulative energy loss of bus $j$ during the period $\Omega_T$ in scenario $s$.

$$\triangle E_{j,s} = \sum_{t \in \Omega_T} \triangle t \cdot \triangle P_{j,t,s}, s \in \Omega_S \quad (38)$$

There might be some repetitive stochastic scenarios among these scenarios generated, whose unserved energy $\triangle E_s$ is close to each other or the same. For example, the unserved energy of scenario $s_1$ and scenario $s_2$ in Fig. 3 is the same. Then, it is expected that $\triangle E_{s_1}$ and $\triangle E_{s_2}$ will also be the same if the DG units with the same size are installed on the same buses. Therefore, it is feasible to use either scenario $s_1$ or scenario $s_2$ to represent both scenarios.

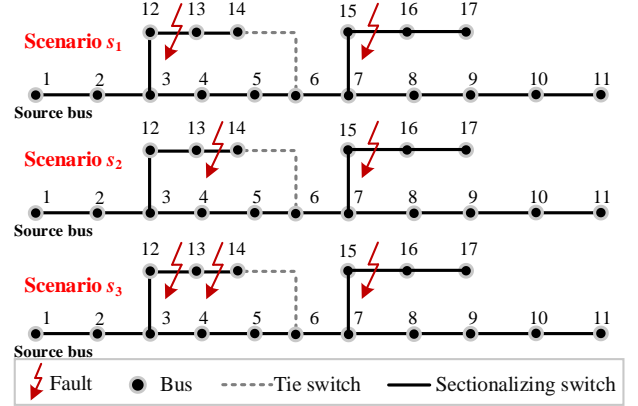

Fig. 3. Example of fault scenarios.

The different number of tripped lines makes a difference to the final strategy. For instance, there is a significant difference between scenario $s_2$ and scenario $s_3$ in Fig. 3. The increase in the number of tripped lines leads to an additional island in scenario $s_3$, significantly impacting the installation strategy of DG units. Therefore, the unserved energy $\triangle E_s$ and the number of tripped lines $\xi_s$ are regarded as the observation $\varphi_s$ for the K-means clustering algorithm. K-means clustering algorithm partitions $S$ observations into $k(k \leq S)$ sets $\{\psi_1, \psi_2, \ldots, \psi_k\}$ to minimize the within-cluster sum of squares [26].

$$\varphi_s = (\triangle E_{1,s}, \triangle E_{2,s}, \cdots, \triangle E_{N,s}, \xi_s), s \in \Omega_S \quad (39)$$

$$\sigma(k) = \arg\min \sum_{i=1}^{k} \sum_{\varphi \in \psi_k} \|\varphi - \mu_k\|^2$$
$$= \arg\min \sum_{i=1}^{k} |\psi_k| Var(\psi_k) \quad (40)$$

$$\mu_k = \frac{1}{|\psi_k|} \sum_{\varphi \in \psi_k} \varphi \quad (41)$$

where $\mu_k$ is the mean of points in $\psi_k$. After the clustering, if a set $\psi_k$ contains multiple scenarios, the scenario closest to $\mu_k$ is selected. The probability of this representative scenario is the sum of all scenarios' probability within this cluster. To find the proper value of $k$, increase $k$ gradually and calculate $\sigma(k)$ by (40). Then, plot a graph of $k$ versus $\sigma(k)$ and apply the elbow method to determine the proper value of $k$.

## IV. SOLUTION METHOD

According to Section III, a finite number of scenarios with the corresponding probabilities have been obtained. Therefore, the problem (30) – (34) can be regarded as a deterministic mixed-integer linear programming. However, the installation strategy of DG units varies with different EEC. Probably, there may be no feasible region for this problem if the EEC is too strict. To solve this challenge, a subsidiary variable $\varepsilon$ is introduced into the optimization model, i.e., $e_i - \varepsilon_i \leq e_i^*, \forall i \in \Omega_N$. Correspondingly, the subsidiary variable should be considered in the objective function, ensuring the ELSI of each community is as small as possible. Additionally, since the proposed model aims to reduce the ELSI of low-income communities, the cost coefficient of the subsidiary variable corresponding to low-income communities has been increased. Finally, the objective function of the first stage is converted to (42), and the EEC is converted to (43).

$$\min \sum_{s \in \Omega_S} \rho_s \cdot g(x, s) + \gamma \cdot \varepsilon_{NLH} + 1.5\gamma \cdot \varepsilon_{LH} \quad (42)$$

$$d^T y - \varepsilon \leq E \quad (43)$$

$$\varepsilon \geq 0 \quad (44)$$

where $\gamma$ is the cost coefficient of the subsidiary variable. $\varepsilon_{NLH}$ is a set of subsidiary variables for high- and medium-income communities, $\varepsilon_{LH}$ is a set of subsidiary variables for low-income communities. $\varepsilon = \varepsilon_{NLH} \cup \varepsilon_{LH}$.

Up to this point, the original optimization model can be expressed as the following deterministic mixed-integer linear programming, which can be solved with the help of solvers, i.e., GUROBI or CPLEX.

$$\min \left\{ \sum_{s \in \Omega_S} \rho_s \cdot q^T y + \gamma \cdot \varepsilon_{NLH} + 1.5\gamma \cdot \varepsilon_{LH} : (x, y) \in \Gamma(s) \right\} \quad (45)$$

where $\Gamma(s) = \{(x, y): Ax \leq b, d^T y - \varepsilon \leq E, \varepsilon \geq 0, T(s)x + W(s)y \leq H(s), x \in \mathbb{Z}_+^{n_1} \times \mathbb{R}_+^{n_2 - n_1}, y \in \mathbb{R}_+^N \}$.

## V. CASE STUDIES

The IEEE 33-bus system and IEEE 123-bus system were applied as the test systems to illustrate the effectiveness of the proposed model. Simulation studies were performed on a computer with Intel Core i7-8650U CPU 1.90 GHz, 16 GB memory, MATLAB 2022a was used as the testing environment, and YALMIP and GUROBI 9.5.2 were used as the solving tool. Considering the DG units' size comes at some discrete value, we assumed the increment of DG units' sizes is 100 kW.

### A. IEEE 33-Bus System

*1) System Description:* Fig. 4 shows the topology of the IEEE 33-bus system, which includes 5 tie switches and 32 sectionalizing switches, which can be actual switches already there or potential locations to install a switch. Also, there are 33 load buses, each of which corresponds to a community. Communities at bus $\Omega_{LH} = \{16, 17, 18, 30, 31, 32, 33\}$ are assumed to be low-income communities at the downstream of the distribution feeder. The reason of this assumption is that after an interruption, a load near the substation bus may be restored faster via network restoration or easy repairment due to location. However, low-income communities often live in an environment with poor infrastructure and less possibility to rerouting to an adjacent feeder [8, 19, 27]. Five buses {10, 18, 25, 30, 32} are equipped with static var compensators (SVCs) to regulate the voltage of this system. Node 1 is the distribution substation which is linked with the wholesale market. The parameters of this system are listed in Table I. It is assumed that $\Delta t$ is 1 hour. It is also assumed that all damaged lines will be repaired at the same time, since this is a long-term planning model.

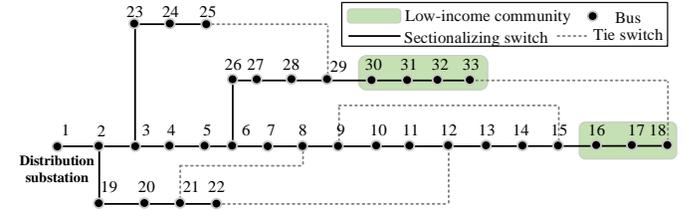

Fig. 4. IEEE 33-bus system.

TABLE I. PARAMETERS OF THE IEEE 33-BUS SYSTEM

| Class | Parameter | Value |
|---|---|---|
| System constraints | Low-income community location | Bus # 16,17,18,30,31,32,33 |
| | $V^{min}$ | 0.95 |
| | $V^{max}$ | 1.05 |
| | $V_{Sub}$ | 1.00 |
| Cost | $\beta$ | $ 50 /kWh |
| | $\gamma$ | $ 100 |
| Investment constraints | $P^{max}$ | 2.5 MW |
| | $N_{DG}$ | 5 |
| | $\alpha^p$ | $ 254/kW |
| | $\alpha^e$ | $ 3.18 \times 10^4$ |
| | $C_{DG}$ | $ 2.0 \times 10^6$ |
| SVC | Location | Bus #10, 18, 25, 30, 32 |
| | Capacity | 0.5 MVar |

*2) Scenario Generation and Reduction Results:* According to the stochastic scenario generation method, 5456 scenarios are generated. Then, the scenario reduction based on the K-means clustering algorithm is performed to achieve the trade-off between computational accuracy and efficiency. Fig. 5 shows the graph of $k$ versus $\sigma(k)$ for the IEEE 33-bus system. Clearly, the decrease of $\sigma(k)$ slows down when $k > 160$. Therefore, $k = 160$ is selected as an appropriate value to optimize the DG units siting and sizing problem.

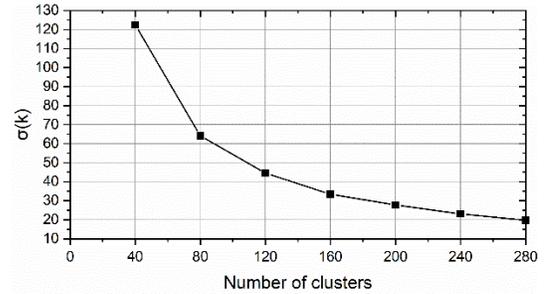

Fig. 5. Graph of $k$ versus $\sigma(k)$ for the IEEE 33-bus system.

*3) DG Units Siting and Sizing Results:* Optimization model (45) with energy equity criterion being 0.02 for all communities ($E = 0.02$) is solved based on selected 160 stochastic scenarios. The optimal strategy is installing DG units on bus $\Omega_G = $

{10, 18, 21, 25, 32}, and with rated power of 2.5 MW, 1.0 MW, 1.6 MW, 1.1 MW, and 1.0 MW, respectively. To verify the effectiveness of installing DG units in enhancing resiliency and achieving energy equity of the power system, the above installation strategies are applied to randomly selected 320 scenarios. The results show that the expected post-fault load shedding is reduced by 87% through installing DG units, greatly increasing the resilience of the distribution system. Also, Fig. 6 shows the percentage reduction in the ELSI and average nodal load of the IEEE 33-bus system. Evidently, there is a relatively larger percentage reduction in the ELSI for communities at DG unit bus 10, bus 21, and bus 25, which have a heavy average load. This is reasonable for reducing the expected post-fault load shedding of the whole power distribution system. Although there is light nodal load for communities at bus 18 and bus 32, two DG units with smaller rated power are installed at these buses to reduce the ELSI for low-income communities and achieve energy equity within this power system. In addition, the percentage reduction in the ELSI shows a relatively flat trend, further verifying the effectiveness of the proposed model in achieving energy equity.

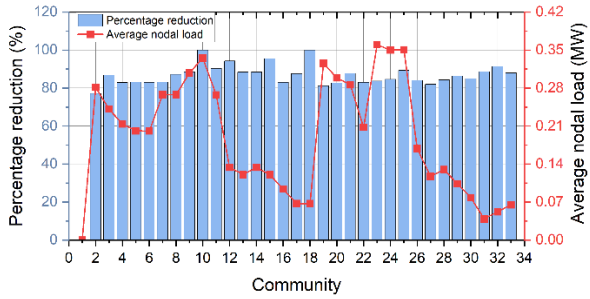

Fig. 6. Percentage reduction in the ELSI and average nodal load of the IEEE 33-bus system compared to case with no DG installations.

*4) Effect of Energy Equity:* The installation strategies of DG units under different EEC are obtained based on selected 160 stochastic scenarios, which are shown in Table II. From Table II, we have the following observations:

**Observation 1**: In general, more DG units are installed at buses with low-income communities under cases with tighter EEC in the resilience-oriented DG planning model.

The above observation is reasonable because the resilience of power systems is improved by utilizing DG units as backup power sources to supply electricity following faults in the proposed model for resilience-oriented DG planning. Therefore, DG units should be installed near the communities of concern to reduce the ELSI following stochastic faults. The EEC in the proposed model ensures that the ELSI for low-income communities remains within acceptable limits. As a result, more DG units should be installed closer to low-income communities under tighter EEC (i.e., smaller value of $E$).

Through implementing these strategies to randomly selected 320 scenarios, the energy equity cost and ELSI of each case are calculated, as shown in Table II and Fig. 7. In Table II, "energy equity cost" is the expected costs of unserved load due to more load shedding, which is on top of the load shedding amount when the EEC is not considered (i.e., $E = \infty$ or the reference case). "Expected costs of unserved load" is the value of objective function in the first stage, i.e., (2). According to Table II and Fig. 7, two additional observations can be summarized:

**Observation 2**: As the EEC is more relaxed (i.e., increasing $E$), both the energy equity cost and its proportion of the expected cost of unserved load is smaller for the resilience-oriented DG planning considering energy equity.

The reason is that a less-binding EEC relaxes the requirement of energy equity (i.e., this may increase load shedding at low-income communities and/or reduce the load shedding at the high- and medium-income communities with heavy loads). Thus, both the energy equity cost and its proportion of the expected cost of unserved load will be lower when $E$ increases. In addition, even if the EEC is the strictest ($E$=0.02), the energy equity cost and its proportion of the expected cost of unserved load is not very high. This alleviates the concerns planners face when implementing resilience-oriented planning with an emphasis on energy equity in practice. Further, it can be observed that when the EEC is sufficiently relaxed (i.e., $E$=0.12), there is no energy equity cost which means the EEC is no longer binding.

**Observation 3**: The ELSI of low-income communities (i.e., loads at buses 16, 17, 18, 30, 31, 32, and 33) is lower in the case with a tighter EEC (i.e., smaller $E$), and the ELSI of high- and medium-income communities is higher but remains at a relatively low level under smaller $E$. As such, the difference in the ELSI between the low-incomes and the high-incomes will be smaller under tighter EEC (i.e., smaller $E$) in the resilience-oriented DG planning.

In summary, a tighter EEC leads to DG units being installed at buses serving low-income communities, which tends to reduce their ELSI. Also, this leads to an increased energy equity cost with respect to the reference case (i.e., no EEC considered). Further, as EEC concerns the load shedding of every community, the ELSIs of the high-incomes and the low-incomes will be maintained at a relatively low level, and more importantly, the ELSI gap between the high-income groups and the low-income groups will be smaller under a tighter EEC. This validates the effectiveness of the proposed method in achieving energy equity.

TABLE II. RESULTS OF THE IEEE 33-BUS SYSTEM UNDER DIFFERENT EEC

| Cases | DG unit bus (#) | Rated power of DG units (MW) | Energy equity cost ($) | Energy equity cost / Expected costs of unserved load |
|---|---|---|---|---|
| $E$=0.02 | 10, 18, 21, 25, 32 | 2.5, 1.0, 1.6, 1.1, 1.0 | 1440 | 28% |
| $E$=0.05 | 10, 18, 21, 24, 32 | 2.3, 0.9, 1.5, 1.7, 0.8 | 840 | 18% |
| $E$=0.08 | 10, 18, 21, 24, 30 | 2.1, 0.9, 1.2, 1.7, 1.3 | 680 | 15% |
| $E$=0.12 | 11, 22, 23, 25, 30 | 1.9, 1.2, 2.4, 0.7, 1.0 | 0 | 0 |
| EEC not considered (i.e., $E = \infty$) | 11, 22, 23, 25, 30 | 1.9, 1.2, 2.4, 0.7, 1.0 | 0 (ref. case) | 0 (ref. case) |

Note: Bus numbers in green indicate low-income community buses.

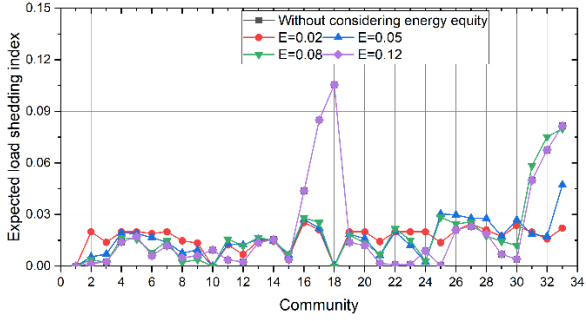

Fig. 7. ELSI of the IEEE 33-bus system under different EEC.

### B. IEEE 123-Bus System

*1) System Description:* The topology of the IEEE 123-bus system, shown in Fig. 8, is employed here to illustrate the scalability of the proposed model. There are 5 tie switches and 116 sectionalizing switches at existing or potential locations. In this system, low-income communities are served by bus $\Omega_{LH} = \{30, 77, 78, 79, 80, 81, 82, 83, 84, 85, 95, 96, 111\}$. Multiple SVCs are installed in this system to provide reactive power compensation to regulate bus voltage. The parameters are listed in Table IV. The parameters of system voltage, cost, and SVCs' capacity are the same as those in Table I. Similarly, $\Delta t$ is 1 hour and we assume all damaged lines will be repaired at the same time.

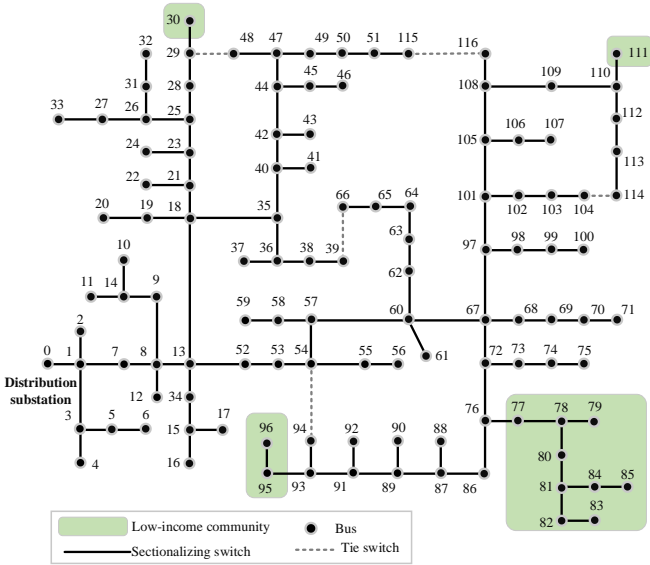

Fig. 8. IEEE 123-bus system.

TABLE III. PARAMETERS OF THE IEEE 123-BUS SYSTEM

| Class | Parameter | Value |
|---|---|---|
| System constraints | Low-income community location | Bus # 30,77,78,79,80,81,82, 83,84,85,95,96,111 |
| Investment constraints | $P^{max}$ | 3.0 MW |
| | $N_{DG}$ | 6 |
| | $C_{DG}$ | $ 4.5 \times 10^6$ |
| SVC | Location | Bus#13,18,27,39,46,50,53,66, 75,93,97,105,108,114,116 |

*2) Scenario Generation and Reduction Results:* Based on the scenario generation method, 7175 stochastic scenarios are generated. The graph of $k$ versus $\sigma(k)$ for the IEEE 123-bus system is shown in Fig. 9. It is evident that $\sigma(k)$ decreases at a much slower pace when the number of clusters is greater than 240. Therefore, 240 stochastic scenarios are selected and applied to obtain the optimal DG unit siting and sizing results.

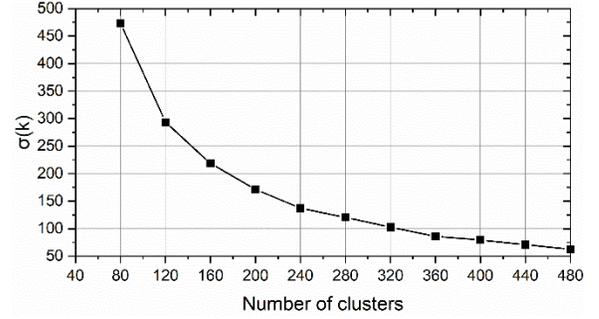

Fig. 9. Graph of $k$ versus $\sigma(k)$ for the IEEE 123-bus system.

*3) DG Units Siting and Sizing Results:* According to the above 240 scenarios, the optimal strategy of the optimization problem with energy equity criterion being 0.02 for all communities ($E$ = 0.02) is obtained, which is to install DG units at bus $\Omega_G = \{30, 51, 77, 95, 96, 111\}$, and with rated power of 3.0 MW, 3.0 MW, 3.0 MW, 3.0 MW, 1.9 MW, 3.0 MW, respectively. Similar to the case studies of the IEEE 33-bus system, 480 scenarios are randomly selected to analyze the effectiveness of the above strategy. The expected post-fault load shedding is reduced by 62%, which shows that installing DG units improves the system resilience. The percentage reduction in the ELSI is calculated and shown in Fig. 10. To improve the resilience of this distribution system, the ELSI of buses with heavy loads is significantly reduced by installing DG units. In addition, there is a light nodal load for low-income communities, but the DG units are also installed on these buses, i.e., bus 30, 77, 95, 96, and 111 to satisfy the EEC. Fig. 10 illustrates that there is a large percentage reduction in the ELSI for low-income communities, achieving energy equity while enhancing the system resilience. There is no change in the ELSI of communities at buses 0-20. Because the upstream branches are the most critical and have a very low probability of outage, those scenarios with damaged upstream branches are not considered in this paper.

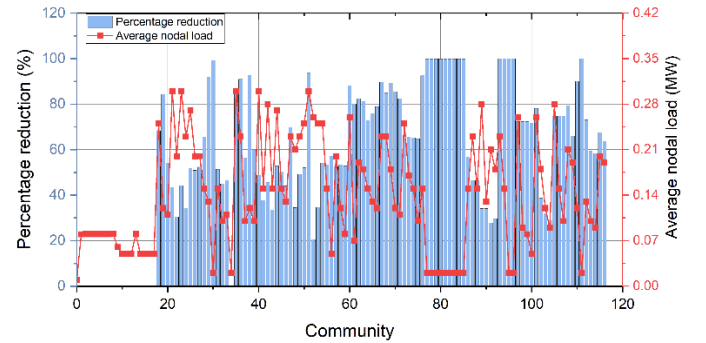

Fig. 10. Percentage reduction in the ELSI and average nodal load of the IEEE 123-bus system.

*4) Effect of Energy Equity:* According to the selected 240 scenarios, installation strategies of DG units are obtained for cases with different EECs, as shown in Table IV. The energy equity costs and ELSI of different cases are calculated based on the selected 480 stochastic test scenarios, which are shown in Table VI and Fig. 11. Similar to the study of the IEEE 33-bus system, we have the same three observations, briefly described below:

- A tighter EEC (i.e., smaller $E$) leads to more DG units installed on the buses with low-income communities.
- A tighter EEC results in a higher energy equity cost and a greater proportion of this cost in the expected cost of unserved load.
- A tighter EEC reduces the gap in the ELSI between the high-income and low-income communities.

These are consistent with the observations from the study of the IEEE 33-bus system and are reasonable. To reduce the expected load shedding for low-income communities and meet the tighter EEC, adequate backup power sources must be provided to low-income communities. Thus, more DG units need to be installed near these communities, reducing the load shedding and their ELSI. Meanwhile, the reduction in ELSI for high- and medium-income communities with heavy loads is less but maintained at an acceptable range. Therefore, to maintain a tighter EEC, we will incur a higher energy equity cost and a greater proportion of this cost in the expected cost of unserved load, with respect to the reference case of no EEC considered. Although the tightest EEC ($E$=0.02) results in the highest energy equity cost and greatest proportion of this cost in the total expected cost, both of them remain in a relatively low level, making it possible to achieve the resilience-oriented planning with energy equity in practice. When the EEC is sufficiently relaxed (i.e., $E$=0.18), there is the same installation strategy as the reference case and no energy equity cost, meaning that the EEC is no longer binding. Further, the ELSI gap between the low-income and the high-income communities will be much reduced with a tighter EEC, which has the goal to reduce the difference of ELSI or resilience for different communities.

TABLE IV. RESULTS OF THE IEEE 123-BUS SYSTEM UNDER DIFFERENT EEC

| Cases | DG unit bus (#) | Rated power of DG units (MW) | Energy equity cost ($) | Energy equity cost / Expected costs of unserved load |
|---|---|---|---|---|
| $E$=0.02 | 30, 51, 77, 95, 96, 111 | 3.0, 3.0, 3.0, 3.0, 1.9, 3.0 | 2860 | 36% |
| $E$=0.07 | 30, 51, 77, 96, 104, 111 | 3.0, 3.0, 2.8, 2.9, 2.8, 2.4 | 1215 | 19% |
| $E$=0.12 | 28, 51, 77, 96, 104, 112 | 3.0, 3.0, 3.0, 3.0, 1.9, 3.0 | 610 | 11% |
| $E$=0.18 | 28, 51, 77, 89, 104, 112 | 3.0, 3.0, 3.0, 3.0, 1.9, 3.0 | 0 | 0 |
| EEC is not considered (i.e., $E = \infty$) | 28, 51, 77, 89, 104, 112 | 3.0, 3.0, 3.0, 3.0, 1.9, 3.0 | 0 (reference case) | 0 (reference case) |

Note: The bus numbers in green indicate low-income community buses.

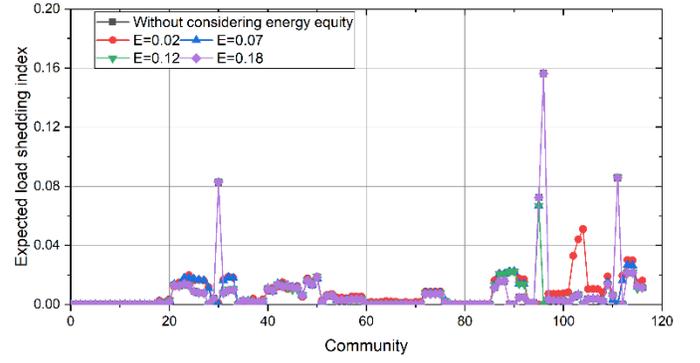

Fig. 11. ELSI of the IEEE 123-bus system under different EEC.

## VI. CONCLUSIONS

In this paper, an actionable method for resilience-oriented DG siting and sizing is proposed to quantitatively incorporate the sociological concept of energy equity into an engineering planning model. The proposed method is based on a two-stage stochastic model to formulate the resilience-oriented DG siting and sizing problem with the EEC, where the energy equity is quantified by the ELSI. Then, a subsidiary variable is introduced to the original mathematical model to ensure it is solvable. Numerical studies are performed on the IEEE 33-bus and 123-bus systems, which verify installing DG units is a valid approach to achieve better energy equity in resilience.

Based on the proposed model and the case studies, we also obtain the following generalized guidelines for DG planning problem with the paradigm of energy equity.

- An effective method to satisfy a stricter EEC is to install DG units in proximity of where the low-income communities are located.
- To achieve better energy equity, there is a higher energy equity cost and a greater proportion of this cost in the expected cost of unserved load, with respect to the reference case of no EEC considered.
- Under a tighter EEC, the difference in the ELSI among different income communities is smaller, and the ELSI of all communities remains at a relatively low level.

The proposed method and corresponding DG planning guidelines may guide the distribution system planners with a practically actionable tool to address energy equity from the perspective of resilience to extreme weather. Therefore, the proposed model bridges the gap between the sociological concept of energy equity and engineering practice in the challenge of DG siting and sizing.